# Smart Asset Management for Electric Utilities: Big Data and Future

**Swasti R. Khuntia[1], Jose L. Rueda[1], Mart A.M.M. van der Meijden[1,2]**

**Abstract** This paper discusses about future challenges in terms of big data and new technologies. Utilities have been collecting data in large amounts but they are hardly utilized because they are huge in amount and also there is uncertainty associated with it. Condition monitoring of assets collects large amounts of data during daily operations. The question arises "*How to extract information from large chunk of data?*" The concept of "*rich data and poor information*" is being challenged by big data analytics with advent of machine learning techniques. Along with technological advancements like Internet of Things (IoT), big data analytics will play an important role for electric utilities. In this paper, challenges are answered by pathways and guidelines to make the current asset management practices smarter for the future.

## 1 Introduction

Asset management, also described as mid-term planning, forms an important activity in the electric transmission and distribution system, the other two being long-term planning (or grid development) and short-term planning (or system operation) (Khuntia, et al., 2016a). Under asset management, maintenance of physical assets and devising maintenance policies form the crucial part. For a recent detailed study on asset management, ref. (Khuntia, et al., 2016b) is recommended. The aims of asset management are to optimize the asset life cycle, improve predictive maintenance and prepare an efficient business plan for investments on new assets. And, with the recent advancements in technology when data and communication play important roles, asset management has evolved from a more traditional to a smarter decision making process. A two-way communication is built, i.e., mechanisms to process the data and the actions taken on the basis of data that may eventually lead to some form of smart function is achieved. Connectivity and data integration are important and necessary as well but not yet sufficient for a smarter asset management. This can be achieved by designing better information management systems that cannot only handle data archiving and retrieval but also help towards data analysis tools.

[1] Dept. of Electrical Sustainable Energy
Delft University of Technology
The Netherlands
e-mail: s.r.khuntia@tudelft.nl

[2] TenneT TSO B.V.
The Netherlands

Using efficient predictive maintenance strategies, utilities can make smarter decisions about when and where maintenance should be performed, which results in reduced maintenance costs with better planning. Moreover, with the advent of advanced computational tools and smart meters, a huge amount of data is collected by the utilities that can be used later for improving the performance of commissioned assets and/or enhancing maintenance policies. Utilizing condition monitoring data for reliability estimation has some similarities with fault diagnosis. The severity of a fault is a key issue for diagnosis, as it presents insight into how long the asset performs before it fails. Identification of the severity of faults is analogous to the estimation of asset reliability. Data mining is useful in this type of study where the aim is to train mathematical failure models and later use them for extended studies like reliability and maintenance optimization. Different data mining techniques serve different purposes, each offering its own advantages and disadvantages. When most of the utilities are trying to follow the condition based preventive maintenance, it is obvious they are moving towards a data centric asset management. Condition-based maintenance of asset health lends itself to large quantities of data. With the use of big data techniques, utilities have to find their way to effectively uncover the patterns, trends, and subtle connections that lay at the root of asset status and component failures. It is important to emphasize that utilities collect large amounts of data every year, but it is not efficiently used because of a number of reasons starting from unavailability of valid mathematical techniques to score of bad/missing data. This work will focus on data collection for asset condition monitoring and the challenges of big data in maintenance including capturing, accessing, and processing information.

The rest of this paper is organized as follows: Section 2 describes the condition monitoring in asset management. Section 3 presents predictive analytics and big data in asset management, how technologies like Internet of Things (IoT) and cloud computing can change the way asset management is studied. This study is concluded in Section 4 with recommendations and discussions on future smart asset management.

## 2   Role of data in asset condition monitoring

Asset condition monitoring includes data collection, condition detection, asset audit, diagnosis, and decision making. It is, therefore, important to focus on collecting right data and improving data quality at the same time. Condition monitor-ing systems are becoming financially attractive, and may even come built-in in the future assets purchased by the utilities (no matter whether they want it or not). However, asset managers and planners still lack straightforward strategies or frameworks to know what information is required in guiding the selection of an appropriate condition monitoring regime. The data requirement for condition monitoring can be divided into two categories: static and dynamic data. Static data refers to asset data that defines the asset itself in normal operating condition. As the name suggests, it does not change during the asset life cycle and is the registered data type often captured. Examples of static data include asset datasheet, location of installation and installation data. Dynamic data refers to asset data that is recorded during the life cycle of asset, mostly operational and maintenance data. This type of data changes during the asset life cycle. Examples are failure rate, maintenance history and other diagnostic data. Due to their volatile nature, dynamic data is often difficult to collect in terms of quantity and quality. So, a rich history of data is often required to perform any kind of maintenance activities. It might be expensive to collect the data as compared to static data, but the advantages overcome the cost.

The interpretation of data coming from condition monitoring systems, the reliability mismatch of diagnostic systems with the equipment being monitored and the volume of data (big data challenge as discussed in section 3) damped the application of condition monitoring systems. Another important issue is the timeliness with which the acquired condition data can be provided and the relationship with the time to failure of this specific asset (CIGRE, 2011). For instance, it is vital to differentiate between condition indicators and health indices though both provide information about asset failure rate. While condition indicators relates to more real-time data indicator, health indices generate information in monthly or yearly time-span. Nowadays, most of these challenges remain and form an obstacle for large scale applications of condition monitoring, especially in combination with the costs for setting up a condition monitoring program.

With ambitious project like GARPUR (GARPUR, 2014), the concept of smart monitoring is introduced which aims to convert condition monitoring of data collected by sensors to visual diagnosis information automatically based on signal processing, signal classification and knowledge rules or expert-based diagnostics systems. To illustrate this, consider the example of partial discharge. Partial discharge is one of the widely used diagnosis methods in power industry to assess cable conditions and can provide an early warning of cable components. It enables maintenance to plan repair and replacement of cables to be carried out timely. For many years, incipient partial discharge faults in power cables have been identified through offline investigation techniques. In the smart grid era, online monitoring will be required for strategically important cable circuits to allow proactive asset management of the cable network to be carried out. Accurate diagnosis of cable conditions, till date, relies on knowledge rules which are based on intensive analy-sis by human experts. This expert knowledge has, so far, been exclusively obtained from offline partial discharge tests. The data acquired, and the rules derived, are often not applicable to online partial discharge monitoring because emphasis of off-line and on-line are substantially different (Zhou, et al., 2009). As a result, using online smart monitoring for automatically performing (or deriving or producing or similar) accurate diagnosis can save labor cost and eliminates errors caused by human. However, to diagnose possible faults by partial discharge, there are some challenges discussed in ref. (Peng, et al., 2013), like:

- developing diagnostic techniques to identify partial discharge signals accurately,
- developing pattern recognition techniques to classify fault signals,
- mining large volume of data and transmitting the data from front-end processor to the control center,
- identifying and localizing possible faults when partial discharge activity occurs in cables

The above discussed techniques are essential to enable smart monitoring by recognizing partial discharge and localizing faults in an autonomous manner. Only by the identification of features, condition monitoring can be applied as smart monitoring in asset management decision making. By generating asset health features or also called health indices in a much more timely and automated fashion, utilities will have better visibility into the overall health of their assets. Health indexing helps in bridging the gap between short-term corrective work driven by condition-based maintenance, and longer-term capital planning.

In such cases, utilities have to find methods that will allow them to assess and monitor the condition of the assets in order to prolong their lifetimes (where possible) because they cannot afford (they do not have the resources and capacity) to refurbish/reinvest the whole network at the same time. Advent of new technologies like Internet of Things (IoT) and deep learning in machine learning will bridge the gap between short-term corrective work driven by condition-based maintenance, and longer-term capital planning, as a result of which acceptance of condition monitoring will play an important role. With advances in data mining and machine learning techniques, deep learning is gaining popularity in the field of asset management and condition monitoring. Deep learning, a branch of machine learning, differs from machine learning in many forms, such as large amount of training data equipped with high performance hardware. Larger the data volume, more efficient the process is. It uses a hierarchical approach of determining the most important characteristics to compare. Based on learning multiple layers of neural network structures, deep learning's success in areas of language modelling, speech and image recognition has made researchers to focus on its application in asset management. One of the advantages of deep learning over various machine learning algorithms is its ability to generate new features from limited series of features located in the training dataset. A deep learning algorithm compris-es of two main parts: training and inferring, which is quite similar to neural networks. The training part involves labeling large amount of data from condition monitoring and extracting the right features while inferring part refers to memorizing the right features to make correct conclusions when it faces similar data next time. This is unsupervised learning as compared to supervised learning with machine learning. It helps asset managers in saving significant time on working with big data while achieving concise and reliable analysis results because it facilitates the use of more complex set of features than the traditional machine learning techniques. Deep learning has not been widely applied to asset health management or prognostics health management field though some attempts have been made in the past. Ref. (Tamilselvan & Wang, 2013) applied a deep learning classification method based on deep belief networks to diagnose electric power transformer health states.

[CHECK to modify the steps]Devising a three-steps methodology can help in leveraging IoT and analytics technologies for improving data collection and component maintenance performance:
1. Collect observational data to assess component condition, defining thresholds or rules to initiate actions or notification to enable condition-based maintenance.
2. Analyze historical data, as well as component failure and work order history, to uncover new patterns that can aid predictions of component failure.
3. Leverage component condition using analysis tools to assess the economic, safety, environmental and public relations effects of failures while also analyzing alternative strategies for handling assets (repair, replace, load shifting, run to fail and so forth).

Though getting to the third step may be the ultimate goal, utilities can derive significant benefit from taking even a first step toward leveraging IoT for condition-based maintenance.

## 3   Predictive analytics and big data for utilities

Predictive analytics in predictive maintenance is used to make predictions about unknown future events. In general, analytics refers to way of interpreting data elements and communicating their meanings to the appropriate users. The introduction of data analytics in the form of data mining, statistical modeling and machine learning techniques help in better future prediction. Of course, it is feasible only if high quality data is recorded/produced on its assets. For instance, depending on utilities, data quality can vary a lot, and consolidating the right data can be difficult. This is because utilities often divide responsibilities for assets among departments that have their own information and communication technology (ICT) systems. Separate ICT systems need to be integrated; so all the asset data they generate feed into a single warehouse. These data also need to be structured so they can be used seamlessly in asset-health indexes, asset-criticality assessments, and asset-management decision models. External databases can help fill gaps in an utilities' own data. The amount of data is going to be huge, as it is commonly called "big data". Big data refers to data sets that are so large and complex they are not easily manipulated using the commonly available database tools. Some of the already big data technologies that have been developed or under development are cloud computing, Internet of Things (IoT), granular computing, and quantum computing. IoT can be seen as a major breakthrough towards future improvements in asset management.

## 3.1  Role of Internet of Things (IoT) in future asset management

Electric utilities have to be proactive in the future because (at least some) will have a tremendous job of dealing with an 'old' network while building the new one. In such cases, utilities have to find methods that will allow them to assess and monitor the condition of the assets in order to prolong their lifetimes (where possible) because they cannot afford (they do not have the resources and capacity) to refurbish/reinvest the whole network at the same time. Internet of Things (IoT) paves way for such a transition.

Talking of Internet of Things (IoT), it is not a single technology and rather a concept in which most new things are connected and enabled. It will pave way for big data analysis and performing operations that were impossible to find without IoT. Some of the advantages why IoT would be beneficial for asset managers and planners include:

- **Requires less human interaction and dependency:** IoT controls the data workflow and can work with less human interactions. It will require less humans and cost to manage the workflows as most of the system will be automated.
- **Better maintenance schedule:** IoT can be used to maximize uptime reliability to ensure that orders can be fulfilled quickly. Maintenance makes it easy to schedule downtime, and maintenance can be scheduled around by proper planning thereby avoiding missed deadlines.
- **Easy tracking of assets:** A program starts with a central database that keeps track of work orders, asset conditions and other data needed to make decisions regarding the management of assets. It is very much necessary for the business to know from where the asset is coming in order to cut the cost. With access to asset data where asset managers can track investment on asset purchase and maintenance, IoT will facilitate in easy handling of resources towards maintenance or investment on new assets.

As the IoT concept enables us to obtain a great amount of data to monitor physical assets, there is an increasing demand for determining asset health conditions in a variety of industries. Accurate asset health assessment is one of the key elements which enable predictive maintenance strategy to increase productivity, reduce maintenance costs and mitigate safety risks.

Smart metering devices (e.g., PMUs, WAMs, SCADA devices) are being installed widely by utilities to collect data as data becomes increasingly important in supporting organizational decisions together with the decision making process. In such case, they are now continually generating more data than at any other time. More data, however, does not necessarily mean better information, or more informed decisions. In fact, many are finding it difficult to use the data. Study reveals that more than 70% of generated data is never used (Koronis, 2006), and also suggests that bad data is worse than no data at all. In other cases, at a needed time, the right data is unavailable or in a different format which makes it difficult for analysis purpose. This situation can be termed as "more data and less information". In today's era, it is just not data anymore, rather big data which is fueling up utilities and it needs more research. There are many major obstacles to the development and implementation of big data analytics in electric power systems, namely, the lack of innovative use cases and application proposals that convert big data into valuable operational intelligence. This repeats the above situation of "more data and less information." More challenges are discussed in next section.

With IoT and big data comes data security and governance into picture. The risk of failing to adhere to data privacy and data protection standards or the advent of data security is a big concern. With big data and advent of new technology, data security and piracy needs to be taken care of. Data security has never been much of an issue for utilities because the infrastructure concerned was hard to access and the potential rewards for hackers were low. Situation has changed lately, proof is that the Ukranian power grid was hacked twice in 2016 and 2017 through the data collecting devices that system operators need at control centers (Zetter, 2016). In the future, as smart homes get connected to transmission network via the distribution grid, it will be easier for hackers to access the power network. Hence, adequate cyber security measures are needed to counter attack cyber-attacks. Cyber security is a huge and evolving challenge for the bulk power system. Progress in terms of efficient methodologies is observed in the last decade and more is expected in future (Ericsson, 2010) (Campbell, 2015).

### 3.2  Challenges in developing big data applications

The challenges in developing big data applications for asset management can be divided into two stages. The first stage is to design a flexible system architecture that accommodates and optimizes big data analytic workloads (ICT side). The second one is to develop scalable mathematical tools capable of processing distributed data (statistical analysis side). The challenges can be further simplified into three types, namely:

1. **Data challenge:** Data collected in power systems suffer from three primary issues. 1) They are incomplete in nature. 2) They come from heterogeneous sources and therefore are difficult to merge. 3) Systems update or make their data available at different rates. Heterogeneity in power system data exists because often the data was intended for a specific application and not collected for a holistic purpose.
2. **Computational scalability:** Traditional mathematical methods are not adequate in handling the inherently large size, high-dimensional, distributed data sets in situ. To address high-dimensionality, machine learning, statistics and optimization algorithms such as classification, clustering, sampling, and linear/nonlinear optimization algorithms need to be easily scalable. Alternatively, scalable and flexible dimension reduction techniques are needed to extract latent features and relevant subsets while balancing accuracy and degree of reduction according to user specification. Large scale data sets in power transmission systems tend to be inherently heterogeneous and distributed. Insufficient research on big data analytics system architecture design and advanced mathematics for large amounts (say, zettabytes of data) to find hidden patterns from the data is an existing hurdle for utilities. To efficiently analyze the distributed data, the algorithms need to come to the data rather than moving data set to the algorithm. Figure 1 shows how unstructured big data can be used to extract classification models for appropriate use cases. In general, such methodology models tend to have two separate steps such as feature extraction and prediction. The first step is to extract features that are indicative of failure or degradation from the data. The second step is to build a prediction model to assess or predict the health condition. This two-step approach involves two separate optimization procedures, which often requires the iteration of the two separate procedures until any acceptable result is achieved.

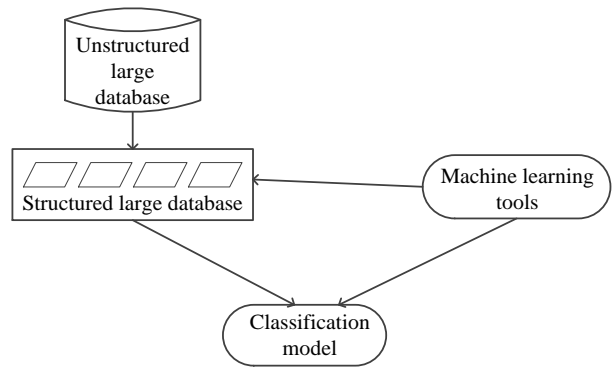

*Figure 1: Supervised learning methodology for big data*

3.  **Making it tractable:** After tackling the scalability issue, the last one is combating tractability. The major classes of data in the asset management decision making process are: 1) failure statistics data, 2) data on the measured conditions of the assets on the transmission network, 3) other data such as weather, load forecasts; 4) asset management data (when and how equipment are maintained and refurbished). Through various types of communication networks, the heterogeneous and complex data sets are transmitted and stored in traditional relational databases, data warehouses, web servers, application servers and file servers. For an efficient use of this data, models should be developed accordingly. Hadoop[2] or Spark[3] is a promising platform since it overcomes many of the constraints with traditional technologies, such as limitations of storage and capacities of computation of huge volume of data. The processing power needed for these platforms is very high, and the cloud computing and distributed technologies are useful tools to be able to analyze data at real-time domain also.

Figure 2 shows the constituents of "big data". The figure can be explained taking an example of the power transformer.

- The *quantity* of data collected and to be collected is going to increase in the future. For instance, an utility collecting data on physical condition of transformers, the quantity would be obviously huge. New data accumulating with the old historical data, and increasing the size of database. There are issues in data capture, transmission, processing, storage, searching, sharing, analysis and visualization. Extracting the relevant data out of gigabytes of data and then transform that relevant data into useful, actionable information is important. Data is increasing at exponential rates, but the improvement of information processing methods is relatively slow which takes us to the next point.
- The *data flow* needs not be constant for all transformers in different locations. As the data flow varies, modelling aspects for transformers will also vary. The condition monitoring is helpful when one can devise maintenance modelling according to the data collected in the database.
- *Accuracy* of data collected is still an open question. Uncertainty due to data inconsistency and incompleteness, latency and model approximations account for accuracy of collected data.
- At the end, *diversity* in the collected data where we recall Figure 1, where it is explained about extracting classification models from structured large databases. In the case of transformers, the health status in terms of gas measurement, visual data like rust and other physical aspects contribute to data diver-sity. While the big data discussion is often centered around quantity, it can be suspected the real problem with data will be diversity. Multiple variables, often from disparate sources such as different sensors on a transformer, can now be used as inputs for inclusion in models which detect the existence of incipient failures. In some cases, the data does not necessarily need to be a numerical value as in rust on the asset's surface. Technological advancements in the future will allow models to read non-numerical values as inputs to predictive maintenance models.

---

[2] Website: http://hadoop.apache.org/
[3] Website: https://spark.apache.org/

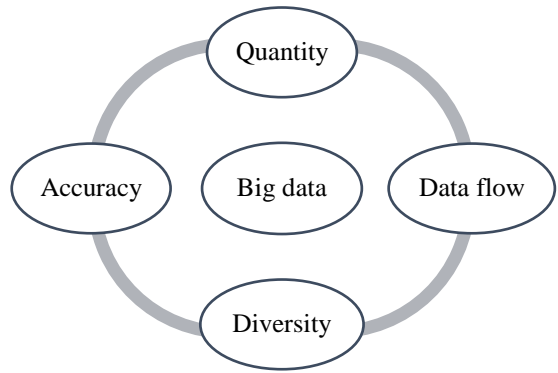

*Figure 2: Components of big data*

A number of questions arises when data collection for condition monitoring is linked with big data. Taking the above example of the transformer, questions on data collection will be like: which part of the transformer is to be monitored, which type of sensor is to be placed and what kind of data are expected from the sensor, how frequently the data is to be collected, whether collected data reached the database on time or not, how to deal with bad data (or missing information). From the huge chunk of data, it will be worth to explore pathways using data analytics to read monitoring data and also learn data correlation (e.g. demand vs weather factors) for efficient maintenance plans. After data collection, new questions on data analytics be like: which technique accurately models the failure rate or ageing, how to deal with missing data and how will it affect the modelling technique, how to deal with situations where there are multiple anomalies detected at the same time. Utilities will need to act soon starting with a small step. Figure 3 shows the pathway in adopting data analytics with four basic steps, namely, building base, learn and adapt, test (small scale), and improvise. Specifying a fixed time-scale for each step is not permissible since different individual steps vary within different utilities. Building a base to collect data while identifying the right tools or database management platforms is vital. Key performance indicators (KPIs) play an important role in data analytics as they determine the significance of each kind of asset data collected. Foundation is not strong without participation of analysts and proper training of existing workforce about data analytics. Once the base is strong, next step is learning and adapting to analytics while enhancing participation in other units within the utility and embracing data security at the same time. Before rolling out the practice in a full-fledged scale, kicking off at small scale is beneficial to utility. Establishing data governance at the application stage is critical. For any utility, data governance can be defined as a collection of practices related to data that ensures security, quality, usability and availability (Khatri & Brown, 2010). Following the small-scale application, adapting to predictive analytics becomes easier with room for continuous improvement

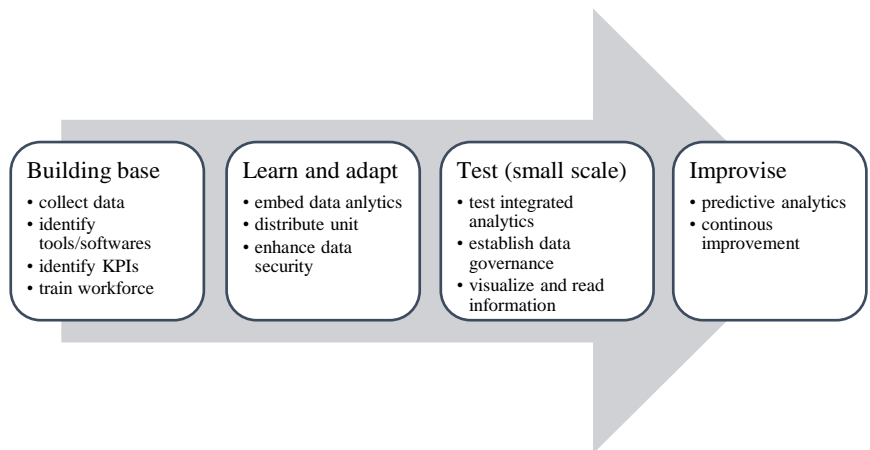

*Figure 3: Future directions on implementing data analytics (KPI refers to Key Performance Indicator)*

## 4   Conclusion

In conclusion, big data will have a large impact on the management of utilities in case of fast deployment of ICT and intelligent sensing within the transmission network. There are many challenges which would affect the success of big data applications in future for utilities. Combination of a significant increase in data, reduced cost of storage, the advancement of cloud based data analytic technology, combined with the emergence of data analyst and scientist roles who know data is the new oil. Presently, experience in integrating big data with current framework is limited. In particular, analytics must be supported by true optimization models to automate the maintenance planning and outage scheduling. It is intended to discover correlations or patterns to make holistic decisions and with the help of analytics utilities can consider all aspects of a decision – the financial side, the maintenance side, as well as the operations side. Also, real application of big data is the ability to understand what data sample is required, ways to analyze and interpret, and then use it. Without completed fields, or validated data, analysis is not possible. So good amount of effort is needed in future to be spent to develop more advanced and efficient algorithms for data analysis that can be easily accepted by utilities. In the end, effective maintenance will be a result of quality, timeliness, accuracy and completeness of information related to machine degradation state, based on which decisions are made.

## Acknowledgement

The research leading to these results has received funding from the European Union Seventh Framework Programme (FP7/2007-2013) under grant agreement No. 608540 GARPUR project http://www.garpur-project.eu. The scientific responsibility rests with the authors.

The authors would like to thank Rémy Clement and Pascal Tournebise of RTE France, Maria Daniela Catrinu-Renstrøm of Statnett SF Norway, and the anonymous reviewers for constructive and insightful feedback.